# HUBS: A dedicated hot circumgalactic medium explorer

W. Cui[a*], J. N. Bregman[b], M. P. Bruijn[c], L.-B. Chen[d], Y. Chen[e], C. Cui[d], T.-T. Fang[f], B. Gao[g], H. Gao[j], J.-R. Gao[c], L. Gottardi[c], K.-X. Gu[d], F.-L. Guo[h], J. Guo[d], C.-L. He[i], P.-F. He[i], J.-W. den Herder[b], Q.-S. Huang[i], F.-J. Li[a], J.-T. Li[b], J.-J. Li[j], L.-Y. Li[g], T.-P. Li[a], W.-B. Li[i], J.-T. Liang[d], Y.-J. Liang[a], G.-Y. Liang[k], Y.-J. Liu[d], Z. Liu[t], Z.-Y. Liu[d], F. Jaeckel[l], L. Ji[m], W. Ji[d], H. Jin[a], X. Kang[u], Y.-X. Ma[d], D. McCammon[l], H.-J. Mo[n], K. Nagayoshi[c], K. Nelms[l], R.-Z. Qi[i], J. Quan[d], M. L. Ridder[c], Z.-X. Shen[i], A. Simionescu[c], E. Taralli[c], Q. D. Wang[n], G.-L. Wang[a], J.-J. Wang[d], K. Wang[i], L. Wang[k], S.-F. Wang[a], S.-J. Wang[j], T.-G. Wang[o], W. Wang[p], X.-Q. Wang[i], Y.-L. Wang[g], Y.-R. Wang[a], Z. Wang[g], Z.-S. Wang[i], N.-Y. Wen[i], M. de Wit[c], S.-F. Wu[q], D. Xu[j], D.-D. Xu[a], H.-G. Xu[r], X.-J. Xue[e], R.-X. Xu[s], Y.-Q. Xue[o], S.-Z. Yi[i], J. Yu[i], L.-W. Yang[d], F. Yuan[h], S. Zhang[t], W. Zhang[p], Z. Zhang[i], Q. Zhong[j], Y. Zhou[a], W.-X. Zhu[d]

[a]Department of Astronomy, Tsinghua University, Beijing 100084, China; [b]Department of Astronomy, University of Michigan, Ann Arbor, MI 48109-1107, USA; [c]SRON Netherlands Institute for Space Research, Sorbonnelaan 2, 3584 CA Utrecht, the Netherlands; [d]Technical Institute of Physics and Chemistry, Chinese Academy of Sciences, Beijing 100090, China; [e]School of Astronomy and Space Science, Nanjing University, Nanjing, Jiangsu 210023, China; [f]Department of Astronomy, Xiamen University, Xiamen, Fujian 361005, China; [g]Shanghai Institute of Microsystem and Information Technology, Chinese Academy of Sciences, Shanghai 200050, China; [h]Shanghai Astronomical Observatory, Chinese Academy of Sciences, Shanghai 200030, China; [i]School of Physical Science and Engineering, Tongji University, Shanghai 200092, China; [j]National Institute of Metrology, Changping District, Beijing 102200, China; [k]National Astronomical Observatories, Chinese Academy of Sciences, Beijing 100101, China; [l]Department of Physics, University of Wisconsin-Madison, Madison, WI 53706-1390, USA; [m]Purple Mountain Observatory, Chinese Academy of Sciences, Nanjing, Jiangsu 210033, China; [n]Department of Astronomy, University of Massachusetts Amherst, Amherst, MA 01003-9305, USA; [o]Department of Astronomy, University of Science and Technology of China, Hefei, Anhui 230026, China; [p]Shanghai Academy of Spaceflight Technology, Shanghai 201109, China; [q]School of Aeronautics and Astronautics, Shanghai JiaoTong University, Shanghai 200240, China; [r]School of Physics and Astronomy, Shanghai JiaoTong University, Shanghai 200240, China; [s]Department of Astronomy, Peking University, Beijing 100871, China; [t]ShanghaiTech University, School of Physical Science and Technology, Shanghai 201210, China; [u]Zhejiang University-Purple Mountain Observatory Joint Research Center for Astronomy, Zhejiang University, Hangzhou, Zhejiang 310027, China

*cui@tsinghua.edu.cn

## ABSTRACT

The Hot Universe Baryon Surveyor (HUBS) mission is proposed to study "missing" baryons in the universe. Unlike dark matter, baryonic matter is made of elements in the periodic table, and can be directly observed through the electromagnetic signals that it produces. Stars contain only a tiny fraction of the baryonic matter known to be present in the universe. Additional baryons are found to be in diffuse (gaseous) form, in or between galaxies, but a significant fraction has not yet been seen. The latter ("missing" baryons) are thought to be hiding in low-density warm-hot ionized medium (WHIM), based on results from theoretical studies and recent observations, and be distributed in the vicinity of galaxies (i.e., circumgalactic medium) and between galaxies (i.e., intergalactic medium). Such gas would radiate mainly in the soft X-ray band and the emission would be very weak, due to its very low density. HUBS is optimized to detect the X-ray emission from the hot baryons in the circumgalactic medium, and thus fill a void in observational astronomy.

The goal is not only to detect the "missing" baryons, but to characterize their physical and chemical properties, as well as to measure their spatial distribution. The results would establish the boundary conditions for understanding galaxy evolution. Though highly challenging, detecting "missing" baryons in the intergalactic medium could be attempted, perhaps in the outskirts of galaxy clusters, and could shed significant light on the large-scale structures of the universe. The current design of HUBS will be presented, along with the status of technology development.

# 1. INTRODUCTION

Although many questions remain, the composition of the universe is now fairly well measured. At present, the dark energy of unknown origin accounts for about 70% of the energy density of the universe, with the rest being matter. The matter is dominated by dark matter (also of unknown origin), with only about 1/6 being of baryonic nature. Therefore, it is the anisotropy in the distribution of the dark matter that determines the distribution of baryons (and thus galaxies) in the universe.

Observationally, a significant fraction of the baryons is still not "seen", constituting a long-standing "missing baryon problem" [1,2]. Theoretical simulations have shed significant light on the issue [3,4,5]: a significant fraction of cosmic baryons might be heated to temperatures of around $10^6$ K by shocks that are produced during the formation of large-scale structures (also known as the cosmic web) and they would emit very weak soft X-rays which are difficult to detect with the present technologies. On smaller scales, for typical galaxies, the fraction of baryons (in matter) seen is much below the cosmological average (~17%). In this sense, baryons are also "missing" from galaxies [6,32]. Observations further suggest that the smaller the mass of a galaxy the lower the baryon fraction. This issue could also be resolved if the "missing" baryons reside in a hot halo surrounding galaxies, whose emission cannot be observed directly at present, due to the lack of observing capabilities in the soft X-ray band.

Physically, baryons do not passively follow the dark matter. As stars in a galaxy evolve, supernovae and the activities associated with the supermassive black hole at the galactic center are known to heat matter in the vicinity to very high temperatures, and might also be able to push the heated matter out to the halo (or even larger distances), forming a reservoir of warm-hot gas. As such gas cools and falls back, it could fuel the formation of the next generation stars in the galaxy. Such cycling of matter in and out of a galaxy is thought to be a key process in its evolution [7], but is still poorly understood, due primarily to the paucity of observations of hot halo gas and also to the complexity of baryonic processes. To further confusing the situation, the galaxy could also accrete warm-hot gas directly from the cosmic web, competing with the processes of baryonic "feedback" that originate from inside the galaxy. One way to distinguish the effects of accretion from those of galactic outflows is to measure the contents of heavy elements (referred to as metallicity in astronomy) in the halo gas, as the accreted gas is expected to be low in heavy elements.

The theoretically-predicted presence of warm-hot gas in the intergalactic medium (IGM) and circumgalactic medium (CGM) has found significant support from indirect observations (including the detection of absorption lines in the spectra of background sources that are associated with foreground warm-hot ionized medium (WHIM)[7,8] and the measured distortion of the spectrum of the cosmic microwave background by the presence of WHIM[9,10]). Although what have been detected likely represent only tip of the iceberg, it does appear that the "missing" baryons are, as predicted, "hiding" in the WHIM. What remain entirely unclear are their detailed properties, such as spatial distribution, temperature, density, and chemical abundances, which is generally considered a major hindrance to understanding galaxy formation and evolution. HUBS is proposed to address this critical observational need[11].

# 2. HUBS DESIGN

## 2.1 Overview

The payload features an X-ray telescope of large field-of-view (FoV) and a non-dispersive spectrometer of high resolution at the focal plane. The spectrometer consists of an array of microcalorimeters[12] that are based on the transition-edge sensor (TES) technology[13]. Trade-offs are considered in the design of the X-ray optics and the matching microcalorimeter array: for a given field of view, the higher the imaging resolution the more pixels are required, but the number of pixels is presently limited by multiplexing TES readout technologies. Taking into account progress in the development of the readout (and other) technologies, we have reached a preliminary design: 60x60 pixel array covering a

1 deg² FoV with modest angular resolution of 1 arcminute. The key design parameters of HUBS are summarized in Table 1. Note that even with a 60x60 array the point spread function (PSF) of the X-ray telescope is still under-sampled.

Table 1. HUBS Specifications.

| Parameter | Value |
| --- | --- |
| Detector Array<br>Regular grid<br>Central subarray | <br>60x60<br>12x12 |
| Energy Resolution [eV] (at 1 keV):<br>Regular array<br>Central subarray | <br><br>2.0<br>0.6 |
| Lower Energy [keV] | 0.1 |
| Higher Energy [keV] | 2.0 |
| Effective Area* [cm²] (at 1 keV) | 500 |
| Field of View [deg²] | 1.0 |
| Grasp [cm² deg²] at 1 keV | 500 |
| Angular Resolution (HPD) [arcmin] | 1.0 |

* The effective area has factored in the solid-angle-averaged throughput of the optics system, filter transmission, and detector quantum efficiency.

The spacecraft is designed by Shanghai Academy of Spaceflight Technology specifically for HUBS. As shown in Fig. 1 (the latest iteration[11]), it is highly integrated with the payload. The X-ray telescope is inside a thermal baffle on the front, focusing X-rays onto the detector through a window (and a series of filters) on the dewar (which sits inside a bore of the spacecraft). The detector and associated readout electronics are cooled by a combination of mechanical cryocoolers and adiabatic demagnetization refrigerator (ADR). Special considerations are made in the design of radiators, to remove heat from the cooling system sufficiently and effectively. HUBS is to be launched into the low-earth orbit around 2030, with a design life of at least five years.

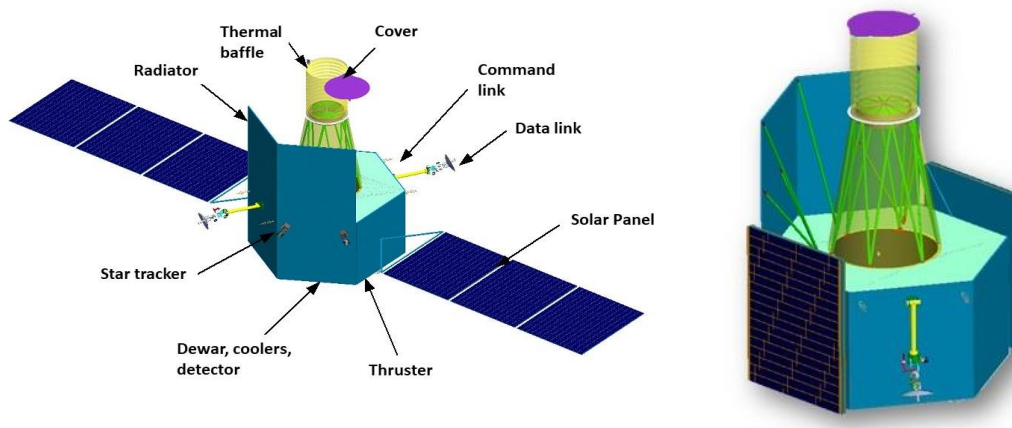

Figure 1. Integrated design of the HUBS spacecraft and payload: (left) with solar panels extended; (right) with solar panels folded. The perspective view of the payload shows the focusing optics assembly, as well as the associated supporting structures, inside the thermal baffle. The dewar is hidden from view.

## 2.2 Design Optimization

*2.2.1 Spectral Resolution*

Achieving high spectral resolution is critical to achieving the primary scientific objectives of HUBS, both for technical and scientific reasons. Technically, it provides a means of dealing with strong foreground contamination. The foreground is primarily due to the X-ray emission from warm-hot gas in and around the Milky Way and that associated with the exchange of charge between ions in the solar wind and neutral atoms in the interstellar medium or in the exosphere of the Earth. The spectrum of the foreground emission is expected to be quite similar to that of the extragalactic warm-hot gas of interest. Fortunately, the spectrum of an extragalactic source is redshifted (with respect to the foreground emission), due to the expansion of the universe, and is expected to be dominated by spectral lines. Simulations have shown that HUBS needs a spectral resolution of 4 eV or better, in order to separate the weak, redshifted lines in the spectrum of a typical galaxy group from the corresponding strong lines associated with the foreground emission[11].

Scientifically, with an X-ray spectrum of sufficient resolution, one can carry out quantitative plasma diagnostics of the emitting WHIM, to derive such important properties as temperature, density, and metallicity. Furthermore, a hybrid design is chosen for the detector, in which a 12x12 small-pixel subarray is used to replace the 3x3 regular pixels at the center of the array. Fully sampling the PSF, the subarray is further optimized to offer even higher photon energy resolution for absorption-line spectroscopy, which involves observing bright, point-like X-ray sources (mainly, active galactic nuclei) that lie at greater distances than the source of interest. Compared with emission lines, absorption lines offer the potential of probing WHIM of significantly lower density and are, therefore, particularly useful for studying the outer region of CGM and IGM.

2.2.2 Angular Resolution

The higher the angular resolution the better HUBS can characterize the spatial distribution of galactic halos (or filamentary structures in the cosmic web). However, as mentioned above, too high an angular resolution might require a detector array with an unrealistic number of TES pixels (as constrained by multiplexing readout technologies). A moderate angular resolution would still be quite useful in minimizing the contribution from background point sources and other interfering sources (e.g., star-forming galaxies) along the line of sight and thus enhancing the signal-to-noise ratio of weak signal detection.

Combined with the high spectral resolution, the design makes it possible to enhance the sensitivity of imaging by focusing on a narrow energy range around bright emission lines. For HUBS, scientifically, the most relevant spectral lines are expected to be hydrogen-like or helium-like oxygen lines around 0.6 keV. One could use narrow-band images to study the spatial distribution of the emitting gas more effectively.

*2.2.3 Energy Band*

HUBS is highly optimized to detect diffuse X-ray emission from hot baryons in the CGM and IGM. Because the X-ray spectrum of WHIM is expected to be dominated by emission lines at energies below 1 keV, we emphasized their detection with the choice of a passing band of 0.1-2 keV for optimizing the design of microcalorimeters. The lower upper bound of the energy range helps reach higher photon energy resolution, because the resolution of a microcalorimeter scales roughly as the square root of its saturation energy.

*2.2.4 Field of View*

For detecting diffuse emission that fills the entire FoV, the sensitivity is determined by the product of the effective area and the FoV (often referred to as grasp), as opposed to the effective area alone for point-like (or compact) sources. For comparison, Athena X-IFU has a grasp of only about 40 $cm^2$ $deg^2$ at 0.6 keV, but has a much larger effective area. HUBS is, in fact, designed to be complementary with Athena in its primary scientific objectives. The large FoV is useful for measuring the spatial profile of extended warm-hot gas around galaxies, galaxy groups, the outskirt of galaxy clusters, and perhaps also filamentary structures in the cosmic web.

*2.2.5 Observing Strategy*

As a dedicated mission for the studies of hot diffuse gas, HUBS will likely spend most time observing low-redshift (nearby) galaxies, galaxy groups, and also galaxy clusters (especially their outskirts). Considering the weakness of the expected signals, the exposure time required is likely quite long (perhaps 1 Ms), therefore, the selection of targets will be very important for achieving the primary scientific objectives of HUBS. In this regard, the portfolio of the HUBS observing program is expected to be limited, compared with general-purpose missions like Athena, but the sensitivity for detecting weak, extended X-ray emission is very high.

**2.3 Design Assessment**

We have carried out an assessment of the scientific capability of HUBS (based on a more conservative design) by making mock observations of galaxies, galaxy groups, and galaxy clusters with data from Illustris-TNG, a state-of-the-art cosmological hydrodynamical simulation. The results show that HUBS would be able to detect the emission lines of He-like and H-like oxygen ions that are associated with warm-hot gas in or around nearby galaxies, groups and clusters, if the targets are selected in the appropriate redshift range, to avoid overlap with the same lines associated with the strong foreground emission or with star-forming galaxies along the line of sight[14]. This highlights the role of target selection in the success of HUBS. Fortunately, there are plenty of sources to choose from, based on known knowledge acquired from observations at other wavelengths. In general, the lower end of the redshift range is determined by the need to avoid foreground lines, and the upper end by the sensitivity of HUBS. We emphasize the oxygen lines because they are expected to be the strongest in the temperature range of interest; the emission lines of other species will also be measured for studying, e.g., the relative metallicity of the gas.

# 3. DEVELOPMENT OF KEY TECHNOLOGIES

**3.1 Detector**

The focal-plane detector of HUBS consists of an array of X-ray microcalorimeters. The microcalorimeter is generally considered as the detector of choice for X-ray astronomy in the future, as high throughput, high-resolution spectroscopy takes on prominent roles. For instance, it has been adopted both by XRISM/Resolve and Athena/X-IFU. It falls in the category of non-dispersive X-ray spectrometer, and is capable of measuring the energy of individual photons with high accuracy and with nearly 100% quantum efficiency (QE). A microcalorimeter consists of three basic components[12]: X-ray absorber, temperature sensor, and a weak thermal link (to the substrate). As a photon hits the absorber, its energy is thermalized, causing the temperature of the detector to increase; as the heat leaks out, the temperature recovers. Therefore, each photon produces a temperature pulse, which is turned into an electrical pulse by the temperature sensor for signal readout electronics. The energy resolution of a microcalorimeter is determined by the accuracy in which the height of the electrical pulse is measured, and is given by

$$\Delta E = \xi \sqrt{\frac{kT^2 C}{\alpha}},$$

where $T$ is the detector temperature, $C$ the heat capacity, $\alpha$ the sensor sensitivity, and $\xi$ a dimensionless parameter of the order of unity. For a given operating temperature, minimizing $C$ and increasing $\alpha$ can, in principle, improve the energy resolution of the detector (although the situation is more complex in reality).

Clearly, the properties of the temperature sensor are the key to the performance of a microcalorimeter. Compared with silicon-based sensing technology (as adopted by XRISM/Resolve), the TES-based technology is expected to deliver higher photon energy resolution, thanks to its higher sensitivity. Moreover, the TES devices are more amenable to multiplexing readout than semiconductor thermistors, making larger arrays possible. A large array of TES-based microcalorimeters can be made with standard microfabrication processes commonly used in the semiconductor industry, and thus promises the potential of high-resolution imaging and spectroscopy with one detector. Like Athena/X-IFU, HUBS also adopts the TES technology for its focal-plane detector. To cover a larger FoV, however, each pixel needs to be larger (about 1 mm on a side). As the absorber defines the size of a microcalorimeter, a serious challenge for the detector development is how to balance large absorbers with the requirement on energy resolution. For comparison, the absorber of a microcalorimeter pixel would be about 16 times smaller in area for Athena X-IFU. A larger absorber is necessarily thinner, to maintain its heat capacity (thus the energy resolution), so mechanical support is an issue. Limiting the passing band to lower energies alleviates the requirement on absorber thickness. Another issue is thermal. The absorber must have excellent thermal conductivity to avoid position-dependent pulse height (and thus degradation of the overall energy resolution); gold is typically chosen for this reason, but it has relatively large heat capacity at low temperatures and is soft mechanically.

For HUBS, the detector development is mainly focused on fabricating Mo/Cu bilayer TES arrays at Tsinghua University and University of Wisconsin-Madison, and Ti/Au bilayer TES arrays at SRON, although the Shanghai

Institute of Microsystem and Information Technology (SIMIT) is also undertaking the development of AlMn alloy and Mo/Au bilayer devices.

At the most advanced stage are the Ti/Au devices developed by the SRON Netherlands Institute for Space Research (SRON). The current TES design is based on rectangular Ti/Au bilayers with a transition temperature $T_c$ of 90 mK. The bilayer is coupled via two stems to a 240x240 $\mu m^2$ Au absorber with a thickness of 2.35 μm. The TESs are fabricated on a 0.5 μm thick SiN membrane to provide thermal isolation from the thermal bath. Further details about the detectors and their fabrication were reported by Nagayoshi et al.[15]. An optical micrograph of a typical TES is shown in Fig. 2.

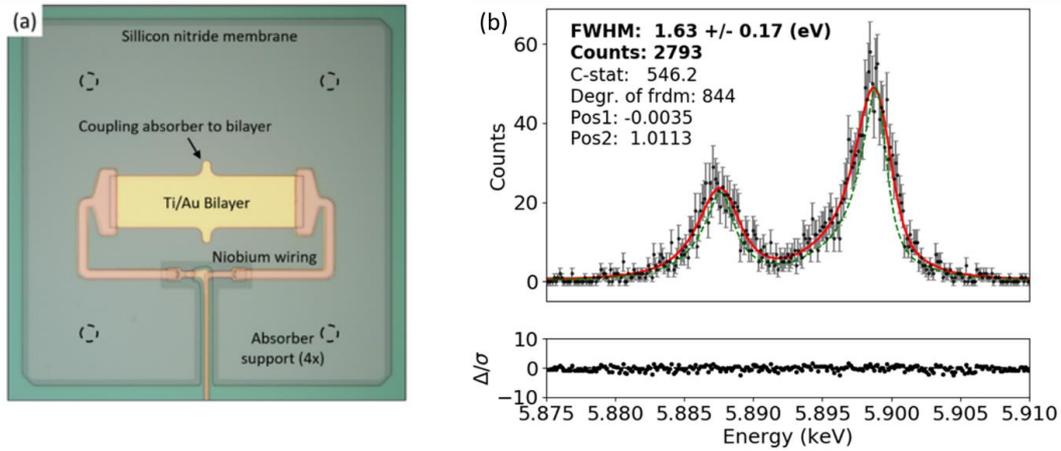

Figure 2. (a) Optical micrograph of a typical SRON TES, labeled with the main components. (b) Energy spectrum of the Mn-Kα line complex. The measured spectrum is shown in black, with the solid red line the best-fit model. The green dotted line is the natural line shape. Note that the TES was AC-biased for the measurement.

In other cases, the processes of making high-quality bilayer and alloy superconducting films are mature[16,17], and small test arrays have been fabricated[16]. A dedicated effort is undertaken at Tsinghua University and SRON on the integration of large absorbers with TES arrays to meet the HUBS requirements, especially electroplating Bi on Au films to improve the absorption efficiency and mechanical properties of absorbers[18].

**3.2 SQUID and Multiplexing Readout Electronics**

Reading out a large TES array operating at < 100 mK requires a multiplexing design[19], to minimize heat load on the cooling system caused by excessive wiring between the cold stage and warm electronics and thus to increase operating time. The time-division multiplexing (TDM) and frequency-division multiplexing (FDM) schemes are most advanced and have seen practical applications. For instance, as shown in Fig. 2(b), a AC-biased TES device was used to measure the energy spectrum of the Mn Kα line complex from a $^{55}$Fe source at SRON (with its FDM setup), and achieved an energy resolution of 1.63 eV[20]. Other possible schemes include code-division multiplexing (CDM) and multiplexing at microwave frequencies (μ-mux). The latter appears to be most promising for even larger arrays beyond kilo-pixels in the future[33].

For HUBS, the development has mainly been on TDM. Critical to TDM are low noise current amplifier and high-speed switch, which are based on superconducting quantum interference devices (SQUIDs) and are being developed at SIMIT. Because a large number of SQUIDs are needed in the TDM readout scheme for HUBS, a reliable fabrication process is required to ensure the quality and the yields of these SQUID devices. SIMIT is developing a new SQUID fabrication process that is based on the standard fabrication process developed in SIMIT for large-scale superconducting integrated circuits using Nb-based Josephson junctions[21]. The process uses an i-line (365 nm) stepper to perform lithography. Nb-junctions with a diameter as small as 1.4 μm can be fabricated reliably with a yield higher than 95%. Fig. 3 shows the cross section of a device fabricated using this process. Using the process, SIMIT has fabricated a SQUID series array containing 22 single SQUIDs. The flux and current noise of this device are 0.56 $\mu\Phi_0$ /√Hz and 10.5 pA/√Hz, respectively. Future design will include electromagnetic simulation results and dummy SQUID to mitigate SQUID resonance and improve the flux coherency.

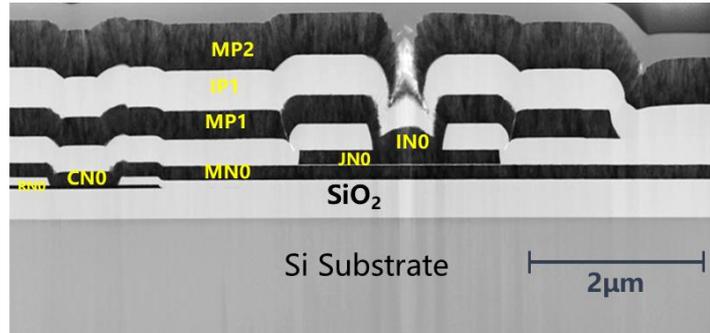

Figure 3. Cross-section of a device fabricated by SIMIT standard fabrication process, courtesy of L.-L. Ying et al.[21]

Superconducting switch is another key element for building TDM circuits. The switch is placed in parallel with the input SQUIDs that are coupled to a TES array. The coupling SQUIDs can be turned on/off by actuating the switch. Usual DC-SQUIDs can be used as flux-actuated switch. SIMIT is developing a similar superconducting switch based on series array of Zappe-style three-junctions SQUID interferometer[22], which has a square-shaped flux-voltage response that can reduce the effect of noise in the control lines.

SQUIDs are also developed at the National Institute of Metrology (NIM). Built upon its experience with large-scale Josephson junction array devices, NIM recently started to develop the Nb/Al-AlOx/Nb-based SIS type of SQUID current sensors for the TES readout. High-quality Nb/Al-AlOx/Nb Josephson junctions were fabricated with the sub-energy gap voltage above 30 mV[23] and $I_c R_n$ value above 1.5 mV. A SQUID based on second-order gradiometric design has been developed. The initial results showed that the flux white noise is about 2 $\mu\Phi_0/\sqrt{Hz}$ at 4.2 K with an input current sensitivity of 17 $\mu A/\Phi_0$[24]. A SQUID array based on 16 first-order gradiometric SQUID cells has also been developed. It shows a current sensitivity of 23.9 $\mu A/\Phi_0$ and the white noise of about 1.2 $\mu\Phi_0/\sqrt{Hz}$. New SQUID designs are under development to improve the noise performance and coupling. NIM is also developing a two-junction superconducting switch for TDM applications.

### 3.3 Cryocoolers and Adiabatic Demagnetization Refrigerator

The HUBS cooling system[25] consists of a precooling stage based on mechanical cryocoolers and a cold stage based on adiabatic demagnetization refrigerator (ADR), as shown schematically in Fig. 4. The precooling system is developed at the Technical Institute of Physics and Chemistry (TIPC), while the ADR at Tsinghua University.

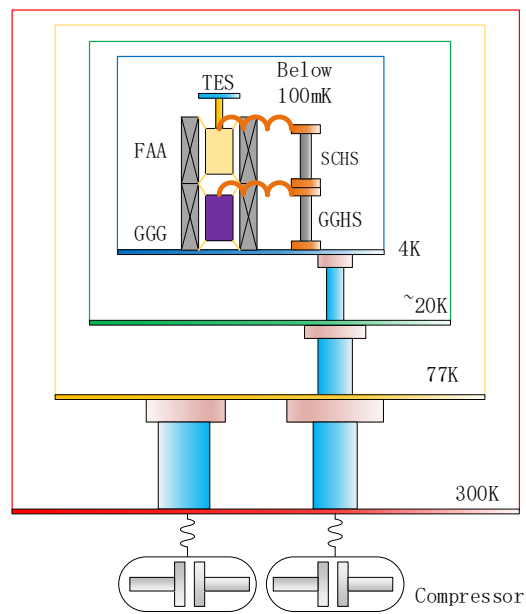

Figure 4. Schematic diagram of the HUBS cooling system. A series of radiation shields are employed at intermediate temperature stages to cut down radiation load from warmer stages.

A high-frequency pulse tube cryocooler (HFPTC) is baselined for the precooling stage[26]. It has the advantages of low vibration and high reliability. A prototype of the thermally-coupled multi-stage HFPTC has been made, as shown in Fig. 5. Using $^4$He gas as working medium, it has reached a temperature as low as 3.47 K (with no load), providing a cooling power of 6 mW at 4.2 K (with an input power of 385 W)[27]. If $^4$He gas is replaced by $^3$He gas, the system would be expected to reach 2.5 K, also with increased cooling power. At present, TIPC is also studying the performance of a gas-coupled multi-stage HFPTC, which would reduce the number of compressors to one (compared with two for the thermally-coupled system). The recent measurements show that with an input power of 450 W the system has reached the temperature 4.3 K, providing a cooling power of 40 mW. Further optimization of the system is being carried out[28].

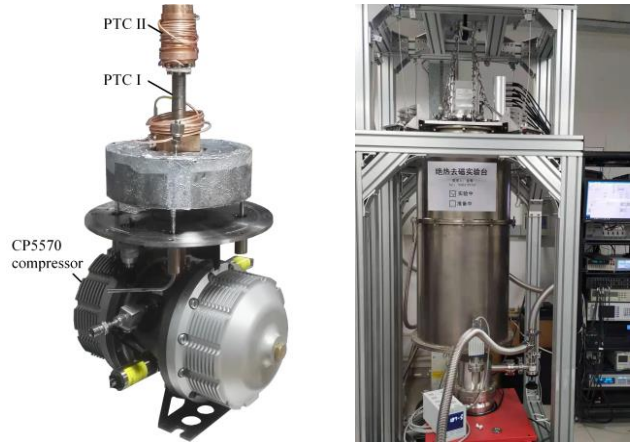

Figure 5. Photos of actual systems: (*left*) a prototype HFPTC developed at TIPC; (*right*) a two-stage ADR prototype (hidden inside the dewar) built and tested at Tsinghua University.

In parallel, a system consisting of two-stage pulse tubes and a Joule-Thomson (JT) junction is being developed as an alternative to the baseline design[29]. The pulse-tube stages employ thermally-coupled HFPTC, with the first stage reaching a temperature of about 80 K and the second stage about 15 K, to enable the JT cooler. As in the baseline design, the pulse tubes cool the radiation shields that are installed at intermediate temperature stages, in order to cut down radiation load on the cold stage where the detector sits. A prototype of the JT system has been made. The latest measurement shows that with an input power of 420 W the system can reach 4.32 K and provide a cooling power of 34.5 mW.

At the cold stage, a two-stage ADR is used to reach the operating temperature of the TES microcalorimeter array (< 100 mK)[30]. The ADR system consists of a superconducting magnet (which can provide a peak field of 4 T), a gadolinium gallium garnet (GGG) paramagnetic salt pill, and a ferric ammonium alum (FAA) paramagnetic salt pill. The GGG salt pill is used to provide a guard stage at about 1 K; the cold plate is further cooled by the FAA salt pill to reach 50 mK. A prototype system was made and tested at Tsinghua University (as shown in Fig. 5). For HUBS, the ADR is designed to provide a cooling power of at least 1 μW (at 50 mK) and a hold time of about 24 hours.

### 3.4 X-ray Optics

The X-ray optics for HUBS is based on the conventional Wolter-I design, with a small focal ratio to realize large FoV and a large number of nested shells to achieve a large collecting area[31]. Studies of true Wolter-I optics and conical approximations are being conducted at Tongji University. The results show that, with optimization, both types could provide the desired FoV and angular resolution. To simplify the manufacturing process, we have adopted a three-stage conic-approximation design for HUBS, as shown in Fig. 6.

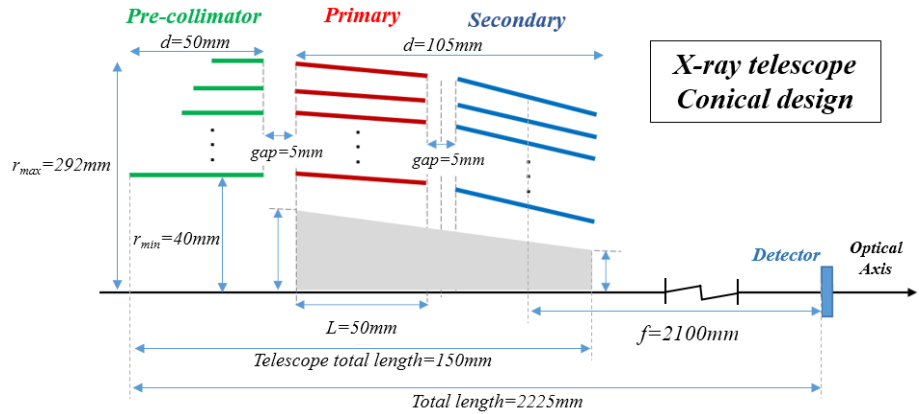

Figure 6. Preliminary design of the HUBS telescope.

In the design, a pre-collimator is placed in front of each mirror entrance to reject stray X-rays (that experience zero or one reflection), similar to eROSITA. At the off-axis angle of 60 arcmin, it can reduce stray X-rays by 68%; at larger off-axis angles, stray X-rays do not reach the detector. Table 2 shows the details of the design.

Table 2. The detailed design of X-ray optics for HUBS

| Parameter | Value | Parameter | Value |
| --- | --- | --- | --- |
| Nested Shell | 215 | Field of view | 62′ @ 0.6 keV |
| Mirror length | 50 mm | Detector area | $39 \times 39$ mm$^2$ |
| Mirror thickness | 0.3 mm | Angular resolution (HPD) | On-axis: 54″ Off-axis: 40″ |
| Focal length | 2100 mm | Geometry area | 1960 cm$^2$ |
| Aperture | 80~583 mm | Effective area | 1118 cm$^2$ @ 0.6 keV 1010 cm$^2$ @ 1.5 keV |
| Gap | 4 mm | Grazing incidence angle | 0.27°~1.98° |

The mirror shells are made with a slumped glass process developed at Tongji University, making them lightweight and suitable for applications that require only a modest angular resolution. The optics development is at an advanced stage[28]. Fig. 6 shows the key processes in the optics development. At present, cylindrical mirror substrates with an angular resolution of 16″ to 35″ are routinely produced.

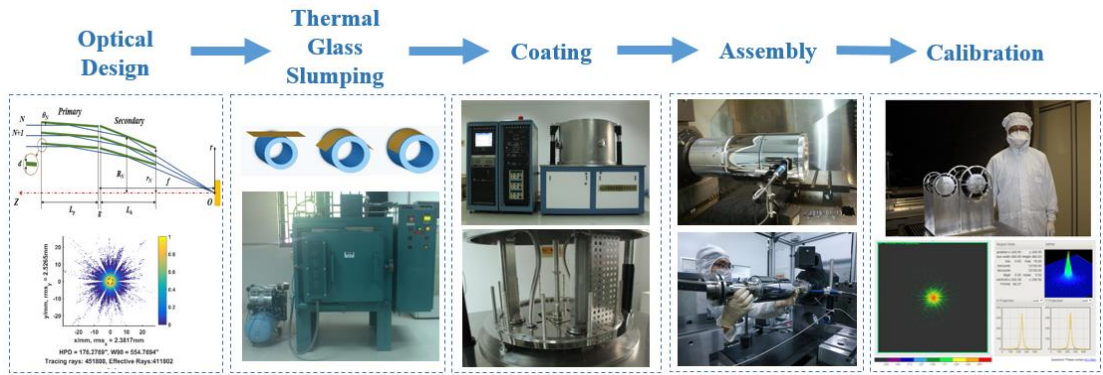

Figure 6. Development of X-ray optics at Tongji University.

A prototype system with 21 pairs of mirror shells was produced and measured at the PANTER X-ray test facility in August 2018, as shown in Fig. 7. The results show that it has an angular resolution of 111″ (HPD) and an effective area of 39 cm$^2$ at 1.49 keV. Another prototype with 3 pairs of mirror shells was made and measured in September 2019. The results show that it has reached an angular resolution of 58″.

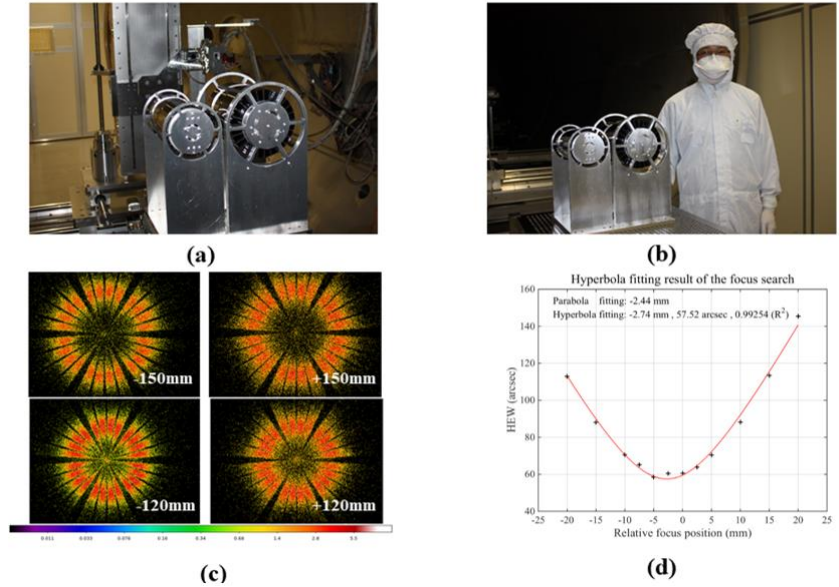

Figure 7. Prototype X-ray telescopes: (top) the photos of 3-layer and 21-layer fully-assembled systems, (bottom) the results of measurements at the PANTER X-ray test facility in Germany.

## 4. SCIENTIFIC ORGANIZATION

Scientific planning for HUBS is organized by the science working groups (SWGs). At present, there are eight SWGs:

- Circumgalactic medium and intergalactic medium
- Galactic feedback processes
- Galaxy groups and clusters
- Simulation and theory
- Supernova remnants
- Active galactic nuclei
- Diffuse X-ray background
- Stars and compact objects

The SWG activities are primarily focused on elucidating scientific objectives of each group, designing observations and selecting suitable targets for achieving the objectives, making mock observations, and optimizing observing strategies. The goal is to have a five-year observing plan for HUBS (with the possibility of a 3-year extension). The HUBS website (http://hubs.phys.tsinghua.edu.cn/en/) contains more detailed information about the mission concept and organization. Scientific workshops and conferences have been regularly held to facilitate collaborations and discussions.

## 5. STATUS UPDATE

HUBS was formally approved in 2018 for a three-year concept study by the Chinese Academy of Sciences through the Strategic Priority Research Program, and is near the conclusion of the study. The project is now being considered by the Chinese National Space Administration (CNSA) for inclusion in its strategic plan for space science over the next national Five-Year Plan period (2021-2025). If selected, the effort will be on key technology development, with a goal of advancing the technical readiness level of each of the HUBS key technologies to 5-6. By the end of the five-year period, a formal review would be conducted and assessment made on whether the project is ready to enter the construction phase.

## ACKNOWLEDGEMENTS

The conceptual study and early development of HUBS is supported by the Chinese Academy of Sciences through the Strategic Priority Research Program, Grant No. XDA15010400. The development of Mo/Cu TES at Tsinghua University received additional support from the National Natural Science Foundation of China through Grants 11927805 and 11821303, from the Ministry of Science and Technology of China through the National Key R&D Program of China, Grant 2018YFA0404502, as well as from the university. The development of SQUIDs and readout electronics is supported by the National Natural Science Foundation of China through Grant 11653004 (to SIMIT), and by the Ministry of Science and Technology of China through the National Key R&D Program of China, Grant 2017YFF0206105 (to NIM) and also by NIM through Grant No. AKYZD2012. The development of the HFPTC system at TIPC is supported by the National Nature Science Foundation of China through Grants U1831203, and that of the JT system by the National Nature Science Foundation of China through Grants 51806228 and 51776213 to TIPC and by the Ministry of Science and Technology of China through the National Key R&D Program of China, Grants 2018YFB0504600 and 2018YFB0504603. The development of X-ray optics at Tongji University is supported by the National Natural Science Foundation of China through Grants U1731242 and 11873004. The work at SRON is partly funded by European Space Agency (ESA) under ESA CTP contract ITT AO/1-7947/14/NL/BW and is partly funded by the European Union's Horizon 2020 Programme under the AHEAD (Activities for the High-Energy Astrophysics Domain) project with grant agreement number 654215.